\documentclass[12pt,preprint]{aastex}

\usepackage{graphicx}
\usepackage{multirow}

\begin{document}

\title{The ion-induced charge-exchange X-ray emission of
the Jovian Auroras: Magnetospheric or solar wind origin?}

\author{Yawei Hui and  David R. Schultz}
\affil{Physics Division, Oak Ridge National Laboratory}
\affil{Bldg. 6010, Oak Ridge, TN 37831 USA}
\email{huiy@ornl.gov, schultzd@ornl.gov} 

\author{Vasili A. Kharchenko}
\affil{Physics Department, University of Connecticut}
\affil{2152 Hillside Road, U-3046, Storrs, CT 06269 USA}
\email{kharchenko@phys.uconn.edu}

\author{Phillip C. Stancil}
\affil{Department of Physics and Astronomy and the Center for 
Simulational Physics, University of Georgia}
\affil{Athens, GA 30602 USA}
\email{stancil@physast.uga.edu}

\author{Thomas E. Cravens}
\affil{Department of Physics and Astronomy, University of Kansas}
\affil{1251 Wescoe Hall Drive, Lawrence, KS 66045 USA}
\email{cravens@ku.edu}

\author{Carey M. Lisse}
\affil{Johns Hopkins University Applied Physics Laboratory}
\affil{SD/SRE, MP3-E167, 11100 Johns Hopkins Road, Laurel MD 
20723 USA}
\email{carey.lisse@jhuapl.edu}

\and \author{Alexander Dalgarno}
\affil{Harvard-Smithsonian Center for Astrophysics}
\affil{60 Garden Street, Cambridge, Massachusetts 02138 USA}
\email{adalgarno@cfa.harvard.edu}

\begin{abstract}
A new and more comprehensive model of charge-exchange induced X-ray 
emission, due to ions precipitating into the Jovian atmosphere near 
the poles, has been used to analyze spectral observations made by 
the {\it Chandra X-ray Observatory}. The model includes for the 
first time carbon ions, in addition to the oxygen and sulfur ions 
previously considered, in order to account for possible ion origins 
from both the solar wind and the Jovian magnetosphere. By comparing 
the model spectra with newly reprocessed {\it Chandra } observations, 
we conclude that carbon ion emission provides a negligible contribution, 
suggesting that solar wind ions are not responsible for the observed 
polar X-rays.  In addition, results of the model fits to observations 
support the previously estimated seeding kinetic energies of the 
precipitating ions ($\sim$0.7-2 MeV/u), but infer a different relative 
sulfur to oxygen abundance ratio for these {\it Chandra} observations.
\end{abstract}


\section{Introduction\label{sec:intro}}
Since the first generation of X-ray observatories, Jupiter has been
one of the primary solar system objects of interest.  In fact, 
Jovian X-ray emission was first detected by the {\it Einstein} 
observatory \citep{metzger1983}, then studied with {\it ROSAT} 
\citep{waite1994, gladstone1998}, and two distinct components were 
identified, namely emission from the high-latitude (polar) and 
low-latitude (equatorial) regions.  Various explanations 
of the origin of these emissions were put forth but could not be 
stringently tested due to the poor spatial and spectral 
(${\cal E}/\Delta {\cal E} < 1$) resolution of the early measurements. 
However, recent observations with {\it Chandra} and {\it XMM-Newton} 
have provided unprecedented resolution on the planet 
\citep{gladstone2002, elsner2005, br2004, br2007a, br2007b, 
br2008, bhardwaj2005, bhardwaj2006} and thorough analysis of the 
observational data have confirmed and refined the characteristic difference 
between Jovian polar and disk emission.  It is now 
generally accepted that the disk component is due to the 
scattering and fluorescence of solar X-rays in the atmosphere 
\citep{maurellis2000, br2007b, bhardwaj2005, bhardwaj2006, cravens2006}. 
In contrast, a mounting body of evidence favors the charge-exchange (CX) 
model as the basic explanation of the Jovian X-ray polar auroras.
This model assumes highly energetic heavy ions precipitating into the 
Jovian atmosphere and emitting K- and L-shell photons after cascading 
from their excited states \citep{metzger1983, horanyi1988, waite1994, 
cravens1995, kharchenko1998, liu1999, kharchenko2006, kharchenko2008}. 

However, fundamental questions remain unanswered, such as What is
the detailed mechanism of ion acceleration before emission; What is 
the source of the heavy ions (do they originate from the Jovian 
magnetosphere and are therefore closely coupled with its ion-rich 
satellites, or are they of solar wind origin); and What are the 
composition of the ion fluxes and their initial kinetic energies at 
the top of the Jovian atmosphere \citep{cravens2003}?  Observations 
and interpretive models answering these questions would advance 
our knowledge of the Jovian system and, by extension, of other
planetary atmospheres and environments. Toward these ends, many studies 
have been carried out seeking to explain the observed spectral features
by fitting them with a continuum plus ion emission lines 
\citep[e.g.,][]{elsner2005, br2004, br2007a, br2007b}. In order
to reflect the intrinsic correlations between the intensities of different
X-ray lines induced by individual ions, CX models were developed to produce
synthetic spectra. These spectra could, in turn, be used to infer parameters 
such as relative ion abundances and initial energies required to account for 
the observations. Early CX models were restricted  to oxygen ion 
precipitation \citep{cravens1995, kharchenko1998, liu1999} owing to the 
lack of extensive atomic data required by the models (i.e., energy-dependent, 
state-selective charge exchange cross sections for all ionization stages of 
the ions and transition probabilities for radiative and non-radiative 
de-excitation, etc.). 

Since oxygen ions contributing to the Jovian X-ray auroras could originate
either in the vicinity of Jupiter or from the solar wind, the next level of 
discrimination between these two possible sources was provided by including
sulfur ions because of their great magnetospheric abundance.  This was 
enabled by the advent of large scale atomic collision and structure 
calculations, and recently included in the CX model 
\citep{kharchenko2006, kharchenko2008}.
These results provided an improved fit to the observations and gave 
indications of the range of initial energies of the ions at the top of 
Jupiter's atmosphere. Here, we seek to further constrain the origin and 
characteristics of the X-ray emission through (i) the addition of carbon to 
the model, again enabled by newly available atomic data, (ii) significantly 
increasing the level of detail of the treatment of radiative transitions,
and (iii) using a more sophisticated methodology to fit the synthetic 
spectra to observations through systematic variation of the input 
ion abundances and kinetic energies.

In this Letter, we report the results of this more complete CX model of 
the Jovian auroral X-ray emission. The basic structure and major improvements 
of the model are described in Section~\ref{sec:model}. In Section~\ref{sec:fit}
a description is given of how our results have been fit to {\it Chandra} 
observations with the X-ray spectral fitting software Xspec 
\citep{arnaud1996}.  Finally, discussion and interpretation of the 
X-ray emission modeling results are given in Section~\ref{sec:dis}.

\section{Ion-induced charge-exchange emission model}\label{sec:model}
A description of the basic CX model along with its development has 
been given previously by \citet{kharchenko1998},  \citet{liu1999},
\citet{kharchenko2000}, and \citet{kharchenko2006, kharchenko2008}, 
and in various reviews \citep[e.g.,][]{bhardwaj2000, bhardwaj2007}.  
In brief, we begin with the Monte Carlo (MC) simulation developed by 
\citet{kharchenko2008} to treat the ion precipitation and CX collisions 
during the ion's deceleration through the atmosphere.  Modeling the energy 
loss and capture into the excited states that seed subsequent radiative 
emission involves three collision channels -- charge exchange, target 
ionization, and electron stripping from the ions.  Each channel has an 
associated energy loss contributing to the stopping process, and stripping 
and charge exchange change the ion's charge state.  CX leads to population 
of excited states that result in photon emission or, less frequently, 
decay non-radiatively.  A synthetic spectrum is built up by tracing all 
possible cascading paths after these excited states are populated.

Specifically, we set the total number of ions, $N_{\rm tot}$, to 1000 in the 
MC simulation for each species considered; increasing this number makes 
negligible changes to the final synthetic spectrum. As pointed out by 
\citet{kharchenko2008}, the spectra are not sensitive to the initial 
charge states of the ions for a fixed ion energy at the top of Jupiter's
atmosphere, so we choose singly charged ions (e.g., O$^+$) to begin the 
precipitation process. The energy loss of the ions due to collisions in 
all three channels are as formulated by \citet{kharchenko1998}.  A 
simulation terminates when the ion slows down to the point that further 
X-ray emission is negligible. 

The most significant improvement in the present work is 
the incorporation of carbon ions as a new source of X-ray emission,
in addition to the previously considered oxygen and sulfur, in order 
to introduce an abundant, more distinctly solar wind originating species. 
This is motivated by  {\it Chandra} and {\it XMM-Newton} observations
\citep{gladstone2002,br2008} which appear to overturn the 
early hypothesis that the auroral 
X-ray emission was linked to ions flowing along magnetic field lines from 
the Io plasma torus.  These new observations indicated that the X-ray
emission comes from latitudes higher than those which are magnetically 
connected to flux from the orbital radius of Io. Thus, other sources have
been suggested including from more distant regions of the magnetosphere 
and the solar wind \citep[e.g.,][]{mauk2003, cravens2003}. 

The MC simulation tracks the energy loss and charge state change at each 
collision according to the probabilities established by the relevant
atomic cross sections \citep{schultz2009} and, for each CX collision,
records the charge state and ion energy. At the end of the simulation
the CX number distribution, $N^{\rm CX}(q,E)$, is tabulated, where $q$ and 
$E$ are the charge state and the energy, respectively. The calculated
probabilities for state-selective capture \citep{schultz2009} are then
multiplied by $N^{\rm CX}(q,E)$ to obtain the initial population 
$X(q,E;\gamma)$ where $\gamma$ represents the energy level ($n, \ell$) 
of each ion populated by a CX collision.

The next step in producing a synthetic spectrum is to calculate the radiative
transition matrices, $T(q;\gamma',\gamma)$, which describes the pathway from
any initial state ($\gamma$) to any final state ($\gamma'$), as
described by \citet{kharchenko1998} and \citet{kharchenko2000}. For the present
work, we collected atomic transition data from the Atomic Spectra Database 
\citep{nist}, the Atomic Line List \citep{all204}, and recent results for 
several ion species \citep{johnson2002, kingston2002, nahar2002}. In total, 
we include 960 carbon, 954 oxygen, and 1512 sulfur emission lines when 
calculating the transition matrices (but in the subsequent model fit to the 
X-ray spectra we only consider lines with photon energy above 200~eV).  This
is an elaboration of the model beyond our earlier work \citep{kharchenko2008} 
which used energy-independent photon yield tables and chose only the most 
prominent lines of oxygen (12 lines) and sulfur (27 lines).  Given the
transition matrices, the photon yields from the cascade process are determined
by $Y(q,E;\gamma',\gamma) = T(q;\gamma',\gamma)\times X(q,E;\gamma)$ and
the synthetic spectrum is found by summing over all ions and charge states.  
Figure \ref{fig:synspec} shows an example of a synthetic spectra for the 
case of equal C, O, and S elemental abundances, initial ion charges equal 
to one, and initial ion energies equal to 2 MeV/u.

\section{Spectral Fitting}\label{sec:fit}
With this CX model, we can seek a fit to the observations by
varying its inputs as in previous work and judging the quality of the 
resulting comparison.  However, in the present case, with three chemical 
species and their differing initial energies, significantly more 
combinations of the inputs need to be evaluated and differences in 
the quality of fits can be made quantitative rather than qualitative.  
For example, with the restriction to oxygen ions in the earliest models, 
given the best available atomic collision and transition data, the only 
input to the model was the initial oxygen ion energy.  When sulfur 
atomic data was first computed and compiled the inputs were the initial
oxygen and sulfur ion energies and their relative abundance. Here, we
construct an Xspec model that has six independent parameters.  These are the 
trial values of the initial energies of the C, O, and S ions at the top 
of the Jovian atmosphere, the relative abundances of C and S with respect 
to that of O (A$_{\rm C}$, A$_{\rm S}$), and the emission line width 
(the full width at half maximum, FWHM, a free parameter of the fitting
procedure, possibly varying from the observational line width).  Because 
we calculate the relative intensities of the emission lines in the CX model, 
the synthetic spectra intensities are given in arbitrary units and the 
normalization (``NORM'' in Tables 1 and 2 below) required by Xspec 
absorbs all other factors affecting the flux. 

We have applied this analysis to the {\it Chandra} observations made by 
\citet{elsner2005}. We first reprocessed the original ACIS-S3 data 
(Observation ID's 3726 and 4418) using the newest calibrations and performed 
all necessary temporal and spatial filtering.  New algorithms were applied 
to adjust the pulse height amplitude values in the original Level 1 event 
files.  The spectra were processed so that each spectral energy bin 
(wavelength bin) had at least ten counts.  All the photons were then 
reprojected into a reference frame fixed on the Jovian disk and the 
spectra were extracted using the latest {\it Chandra} data analysis 
package, ciao 4.0.1. 

We attempted fits of the north and south polar observations 
3726 and 4418 with C, O, and S but found that the carbon ion 
energy of the fits approached the very low value of 0.03 MeV/u.  At 
such energies, the population of highly charged ions producing X-rays 
is negligible.  Closely related to this behavior the fits yielded 
no bound for the carbon to oxygen abundance ratios, meaning that the relative 
amount of carbon required exceeded the preset limit we imposed, in this case 
a C to O ratio of 1000, or dropped to the lower limit, zero.  To illustrate 
this behavior of the fits we display in Table~\ref{tb:w_wo_C} results including 
C and omitting it for spectra from the north polar region in observation 3726.  
The first line of the table shows the best fit value of the carbon energy 
($E_{\rm C}$) being very low and the relative abundance of carbon at 
the preset limit value of 1000. The reduced chi-square value of the fit and 
the degree of freedom (DOF) of the fit - a measure of the signal to noise 
ratio of the fitted spectrum - indicate that the fit is reasonably good.  
However, since the carbon energy is so low and the relative abundance 
has to be so high to have an effect, this fit indicates that carbon 
could be neglected.  In fact, attempting the fit without carbon, as 
shown in the second line of Table ~\ref{tb:w_wo_C}, yields a comparably 
good fit and does not greatly change the O and S initial energies, the 
S to O abundance ratio, or the FWHM of the lines. (We note that the 
fits always favored line widths of around 55 eV, so after the initial 
fitting we have fixed this input variable to accelerate the calculations.) 
Therefore, our results indicate a negligible contribution of carbon to this 
spectrum.  As we will report in a subsequent detailed report, we have also 
analyzed the {\it XMM-Newton} observations made by \citet{br2004,br2007a} 
and reach a similar conclusion.

\begin{figure}
\mbox{
\includegraphics[angle=0,scale=0.32]{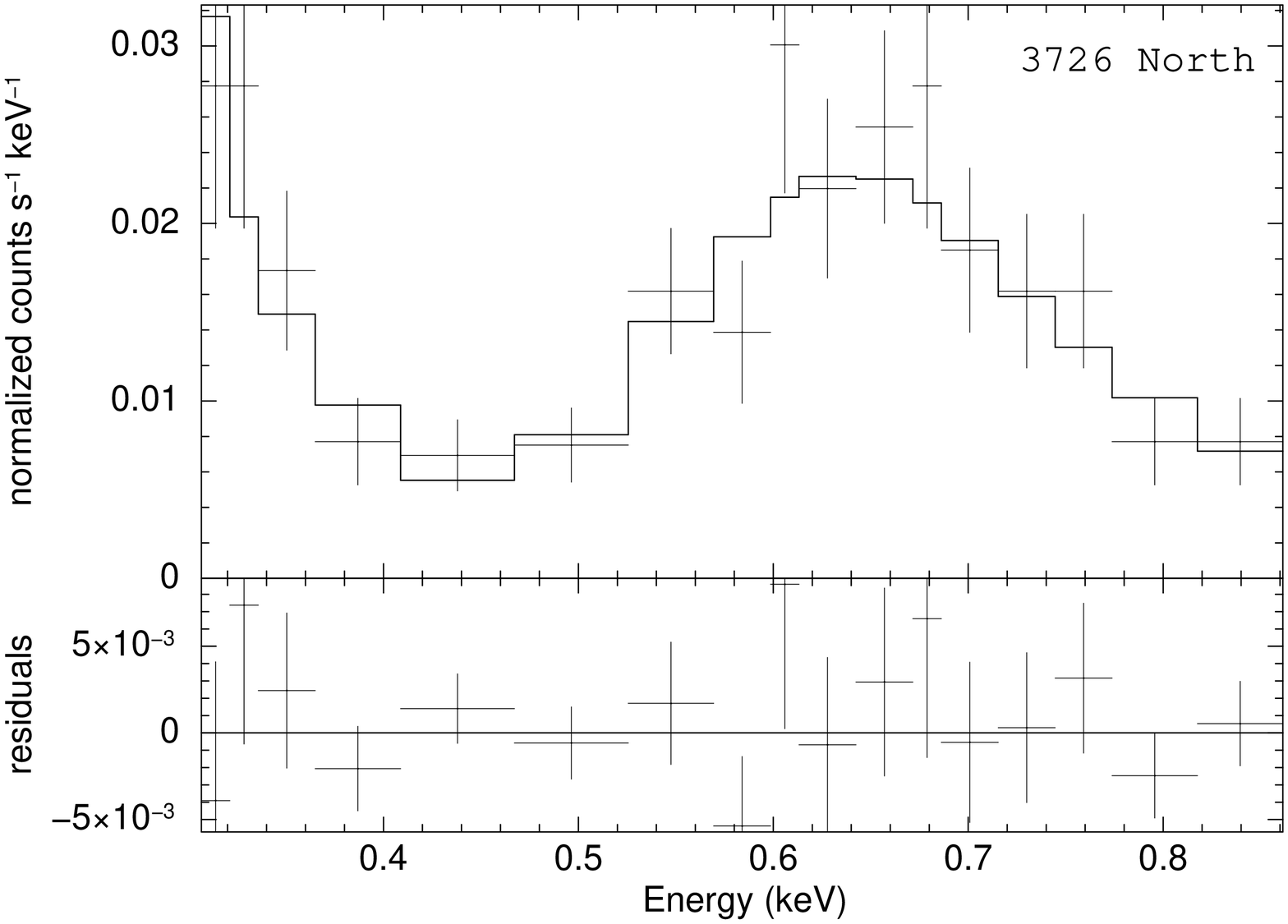}
\includegraphics[angle=0,scale=0.32]{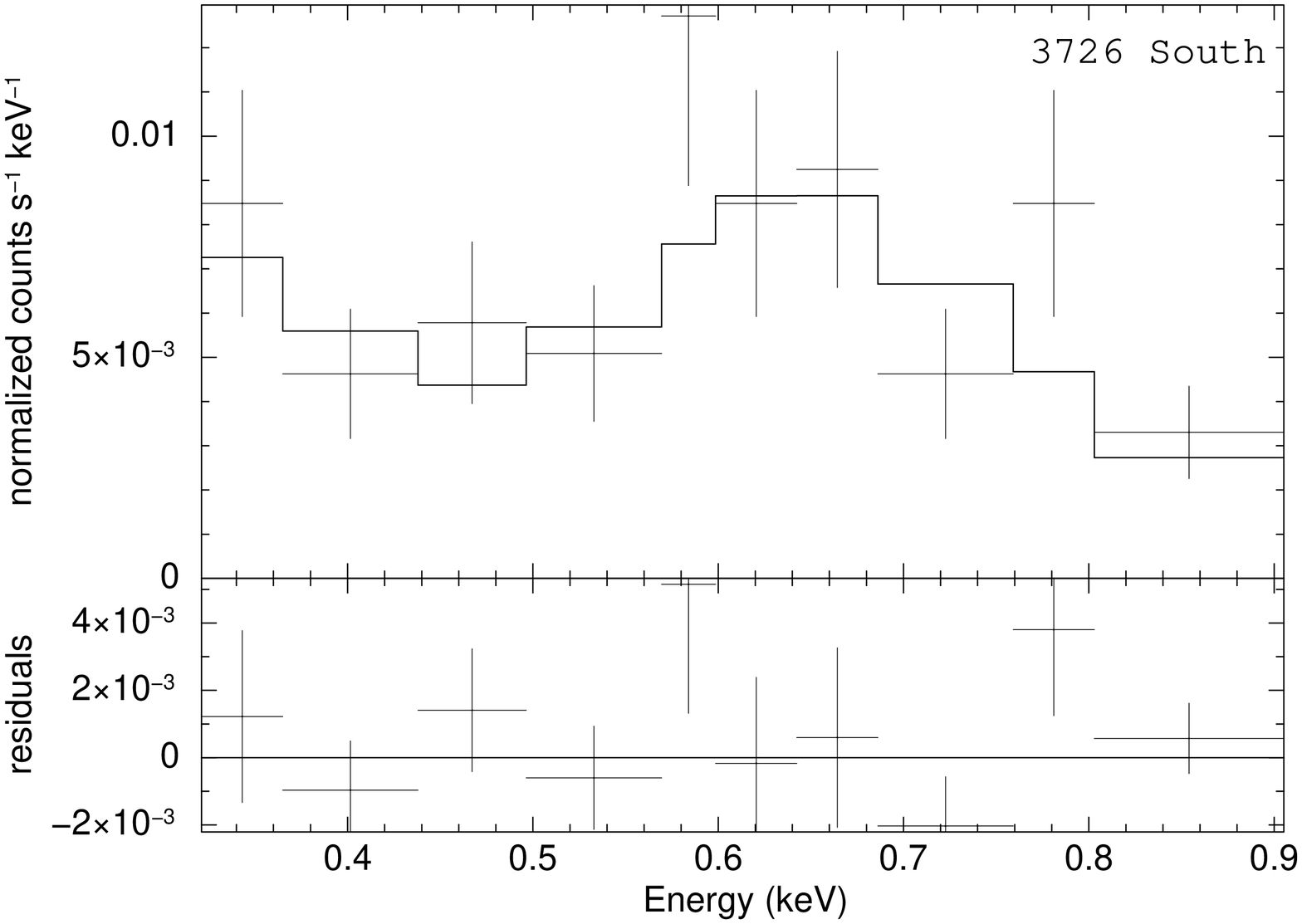}
}
\mbox{
\includegraphics[angle=0,scale=0.32]{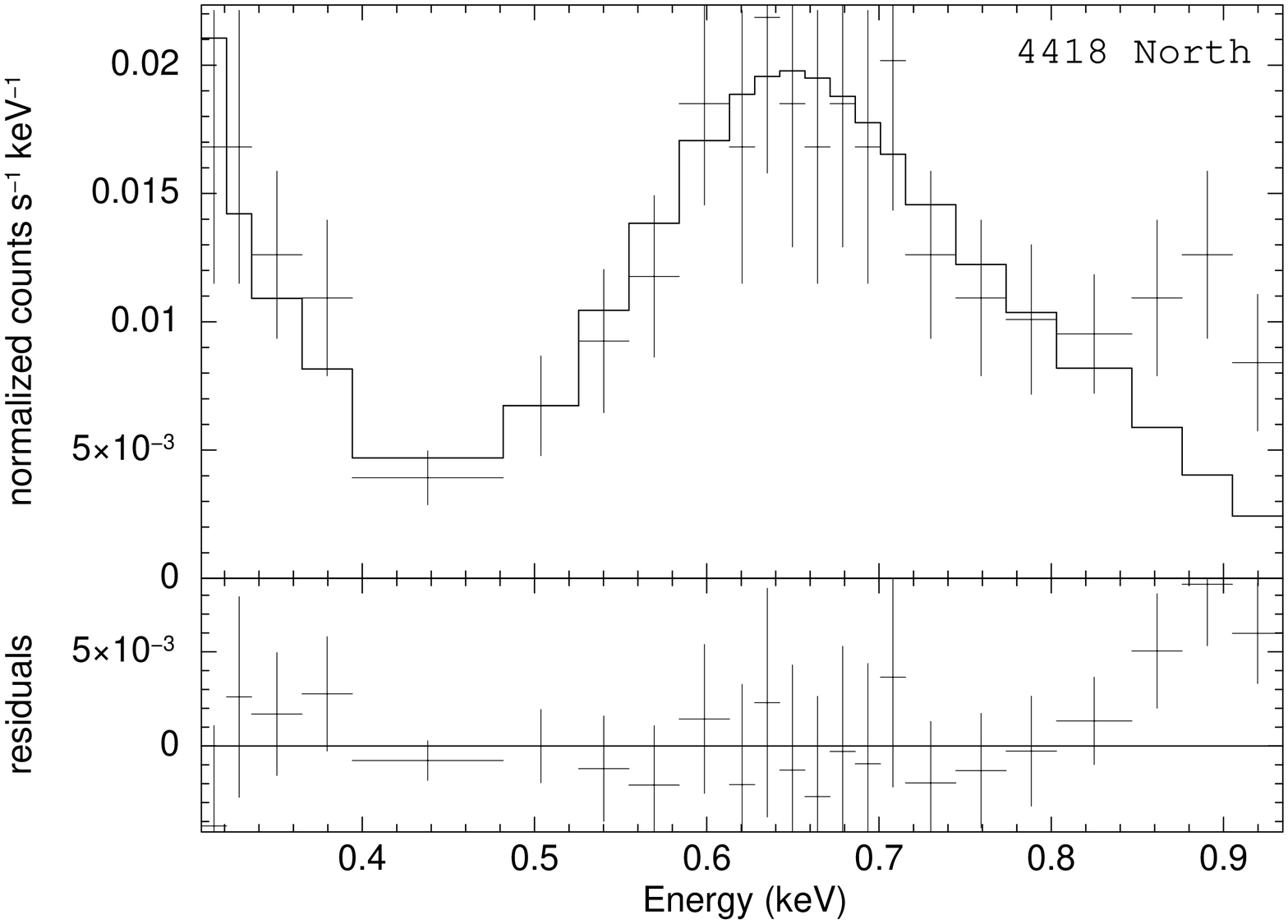}
\includegraphics[angle=0,scale=0.32]{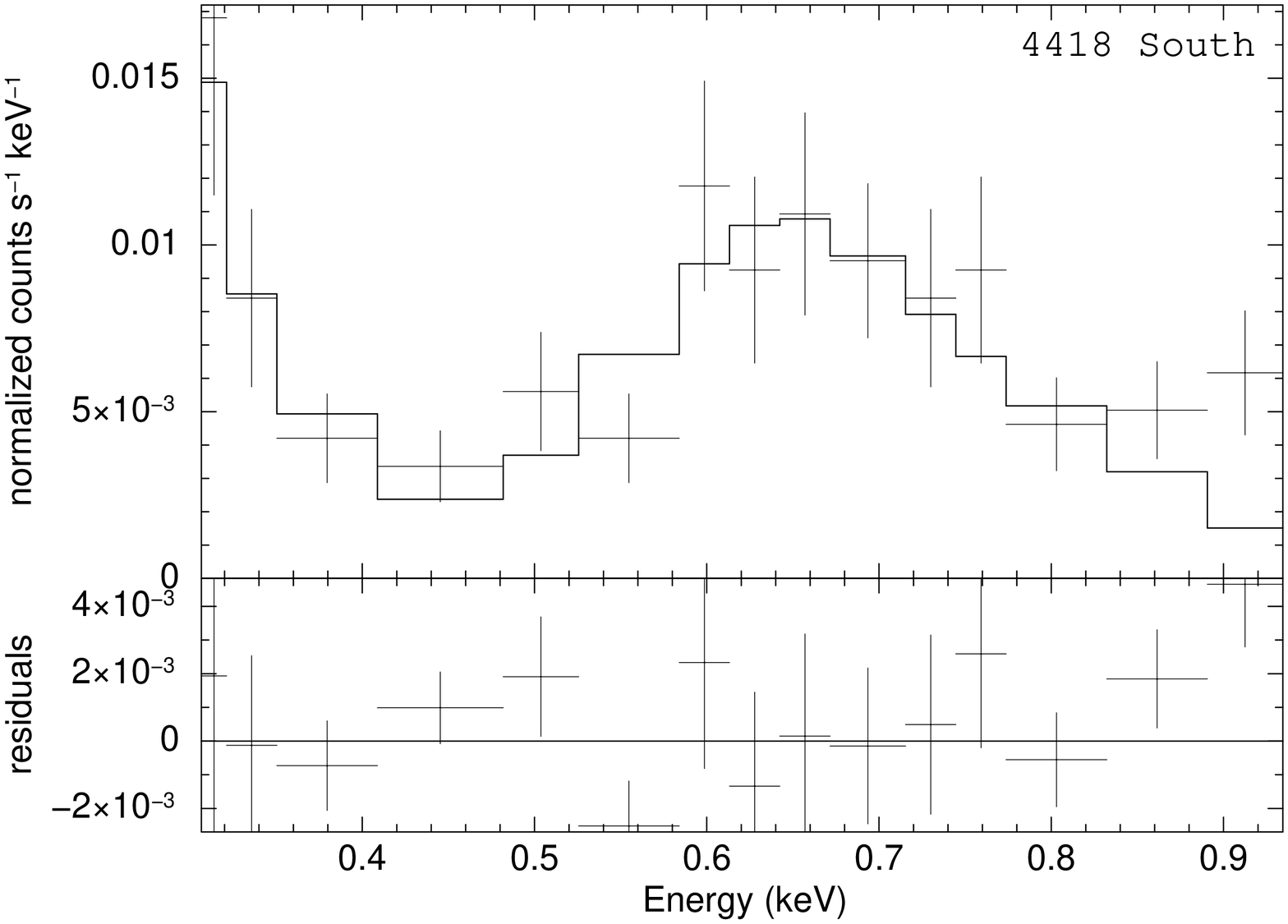}
}
\caption{Fits to the spectra of the north and south polar X-ray auroras 
observed by {\it Chandra} (observation IDs 3726 and 4418) neglecting carbon 
in the CX model.  The fit parameters are given in Table 2.
\label{fig:fits}}
\end{figure}

In Table~\ref{tb:chanfit} and Figure~\ref{fig:fits} we show the model fits to 
both the north and south polar auroral spectra ({\it Chandra} observation IDs 
3726 and 4418) neglecting carbon. The inferred oxygen energy is near or at the
largest value included in our CX atomic dataset and for sulfur it is within
the range of about 0.7 to 1.5 MeV/u, both in reasonable agreement with previous
conclusions \citep{kharchenko2008}. The fits favor a sulfur to oxygen ratio
greater than one, but we find that this is sensitive to the spectral resolution
and our preliminary fits to {\it XMM-Newton} observations indicate smaller 
values of this ratio.

\section{Discussion}\label{sec:dis}
If the solar wind played an important role as a source of ions leading to 
the CX driven X-ray emission near the Jovian poles, then carbon should be
required to account for the observed spectra, because apart from the 
dominant components of the solar wind, hydrogen and helium that do
not contribute to the X-ray emission, oxygen and carbon are the 
most abundant X-ray producing species \citep{vonS2000}.  Since our model
shows no improvement to the fit to {\it Chandra} observations by including
carbon we conclude that ions of magnetospheric origin are dominant in
driving this X-ray emission. Moreover, in other observations, namely
those detecting X-ray emission from comets \citep{schwadron2000, lisse2001, 
lisse2005, lisse2007, bodewits2007} clear identification of C and O lines
have been made.  In that case, the C and O (as well as N, Mg, Fe, for 
example) are definitely of solar wind origin, implying that if solar
wind C contributed at Jupiter, it would likewise be detected.

\begin{figure}
\hspace{-2.0cm}\includegraphics[scale=0.9]{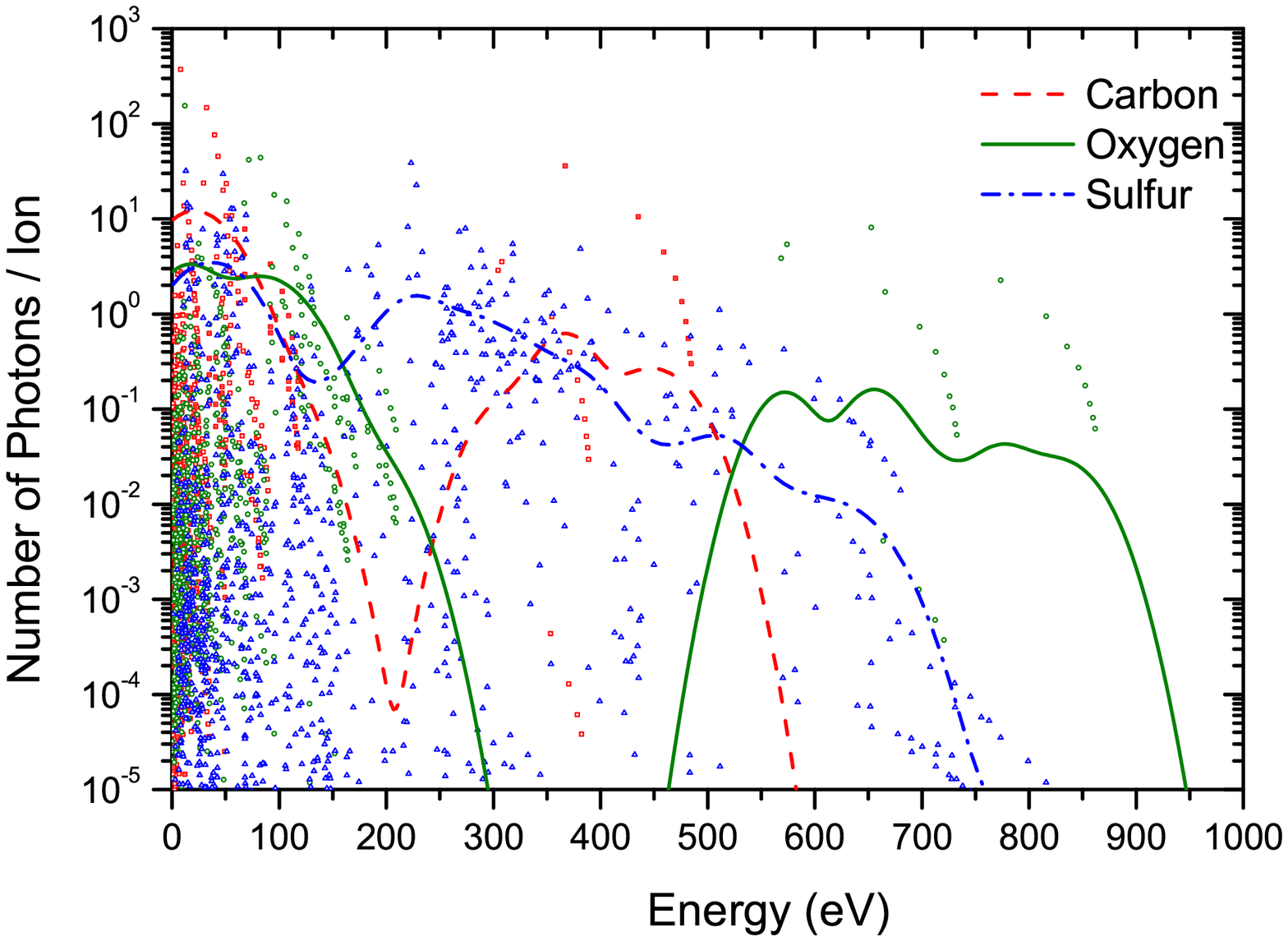}
\caption{Example synthetic spectra for the case of equal C, O, and
S ion abundances and $N_{\rm tot}=1000$, $q_{\rm init}=1$ and 
$E_{\rm init}=2$~MeV/u.  Individual lines are marked with different 
symbols and colors for different elements (carbon: red squares; 
oxygen: green circles; and sulfur: blue triangles).  Also shown 
(smooth lines) are convolutions of these lines with a Gaussian 
line profile of 24.5 eV FWHM and with all lines belong to a single 
element summed. The line width used in this illustration was chosen
to be the same as in previous work \citep{kharchenko2008}.
\label{fig:synspec}}
\end{figure}

This conclusion also implies that solar wind carbon ions are not 
accelerated above 1 MeV/u before precipitation into the atmosphere 
\citep{cravens2003}, whereas a mechanism for magnetospheric
ions accelerates oxygen and sulfur ions to this energy or above.  
For example, as shown in Fig.~\ref{fig:synspec}, the most significant 
flux due to carbon occurs between 300 and 500~eV (with a peak due to
the C$^{5+}$ (2p $\to$ 1s) transition at 367~eV) if given 
sufficiently high initial ion energies and abundance.  However, the low 
carbon initial energy inferred from the model fit makes it unlikely  
that carbon ions participated in enough CX collisions in the high 
ionization stages during the precipitation to contribute appreciably.  
This fact is also reflected by the very large uncertainties accompanying 
the carbon abundance fits. 

It is also interesting to note that the fits favor sulfur ion velocities
(energy per mass, $m$) about half those for oxygen.  This is indicative 
of the original sulfur ion population in the outer magnetosphere
being mainly singly charged ($q$=1) rather than doubly charged or higher.  
That is, as suggested by \citet{cravens2003} it is likely that a 
field-aligned potential accelerates the ions to the energies needed 
to produce X-rays. Then, since the total kinetic energy produced by a 
potential difference is proportional to $q$, the energy per mass unit 
is proportional to $q/m$ of the original ions, suggesting
a dominantly O$^+$ and S$^+$ initial ion population with energies per
mass unit for sulfur ions half those of oxygen ions.

We also note a favoring of a greater abundance of sulfur relative 
to oxygen in the present fits compared to our previous results 
\citep{kharchenko2008}.  This comes about from a rather sensitive dependence 
of the fits on the low photon energy {\it Chandra} spectra.  In particular 
the present best fit sulfur energies are generally smaller than those 
inferred previously, so that the sulfur abundance must be increased to make 
up for the flux at the lower range of X-ray energies since sulfur is the 
dominant source for the spectral feature near 300 eV. However, our 
preliminary fits to similar {\it XMM-Newton} observations show lower
A$_{\rm S}$ for some observation dates, illustrating the 
sensitivity to the lower energy portion of the spectrum and the possible
temporal variability of the ratio. Since a putative source of the magnetospheric
sulfur and oxygen ions would be from SO$_x$ ($x$=1,2,3), further
observations improving the reliability of the lower energy portion of the
spectrum and increasing the coverage of it in time could provide important
constraints on the relative abundance of S and O ions and thus their
origin and the acceleration mechanism.

Thus, additional observations would clearly be of significant help to 
test the CX model and its inputs and thereby improve our understanding
of ion sources, distributions, and acceleration mechanism.  For example,
observations during other points along the solar activity cycle might help 
pick up a changing contribution from the solar wind and more and deeper 
observations would help show the range of the variability of the X-ray 
aurora.  

\acknowledgments{Acknowledgements. This work has been supported by NASA 
Planetary Atmospheres Program Grant NNH07AF12I.  We are also grateful 
to A.~Bhardwaj and G.~Branduardi-Raymont for helpful discussions, and 
to the {\it Chandra} Helpdesk staff for advice on and assistance with 
processing the raw observation files.}

\begin{deluxetable}{l l l l l l l}
\tablewidth{0pt}
\tablecaption{Spectral fitting of the north polar spectra of {\it Chandra} 
observation 3726 with (first line) and without (second line) carbon. The 
columns give best fit initial ion energies (E$_{\rm C}$, E$_{\rm O}$,
E$_{\rm S}$), abundances relative to oxygen
(A$_{\rm C}$, A$_{\rm S}$), line width (FWHM), spectrum normalization (NORM), 
and reduced chi-square with degrees of freedom (DOF, defined as the number of 
spectral energy bins minus the number of independent inputs). 
All uncertainties are computed with 
$\Delta\chi^2=2.706$, equivalent to $90\%$ confidence for a single parameter. 
``$\ast$'' in the errors means that the parameter is not bounded in that 
direction.  The superscript outside of the bracket represents the power of 10.
\label{tb:w_wo_C}}
\tablehead{\colhead{OBS ID} & \colhead{$E_{\rm C}$ (MeV/u)} 
& \colhead{$E_{\rm O}$ (MeV/u)} & \colhead{$E_{\rm S}$ (MeV/u)} 
&\colhead{A$_{\rm C}$} & \colhead{A$_{\rm S}$} }
\startdata
\multirow{2}{*}{North 3726} & $0.03^{+0.26}_{-0.03}$ & $2.00^{+*}_{-0.39}$ 
& $0.81^{+0.05}_{-0.45}$ & $1000^{+*}_{-*}$ & $4.7^{+1.5}_{-2.3}$  \\
& -  & $1.48^{+*}_{-0.28}$ & $0.81^{+0.48}_{-0.45}$ & 0 & $5^{+243}_{-3}$  \\ \\
\hline 
OBS ID & FWHM (eV) & NORM & Red. $\chi^2$ [DOF]  \\ 
\hline \\
\multirow{2}{*}{North 3726} & $56^{+7}_{-44}$ & $(1.1^{+1.2}_{-0.4})^{-6}$ 
& 0.84 [10] & &  \\
& $55$\tablenotemark{\; a} & $(1.06^{+0.52}_{-0.32})^{-6}$ & 
0.64 [13] & &  \\
\enddata
\tablenotetext{a}{FWHM is fixed at 55~eV (see text).}
\end{deluxetable}

\begin{deluxetable}{l l l l l l}
\tablewidth{0pt}
\tablecaption{Spectral fitting of the north and south polar spectra of
{\it Chandra} observations 3726 and 4418 without carbon. As in Table 1
all uncertainties are computed with $\Delta\chi^2=2.706$, equivalent to 
$90\%$ confidence for a single parameter. ``$\ast$'' in the errors means 
that the parameter is not bounded in that direction. 
The superscript outside of the bracket represents the power of 10.
\label{tb:chanfit}}
\tablehead{\colhead{OBS ID} & \colhead{$E_{\rm O}$ (MeV/u)} & \colhead{$E_{\rm S}$ (MeV/u)} &
\colhead{A$_{\rm S}$} & \colhead{NORM} & \colhead{Red. $\chi^2$ [DOF]} }
\startdata
North 3726 & $1.48^{+*}_{-0.28}$ & $0.81^{+0.48}_{-0.45}$ & $5^{+243}_{-3}$ &
$(1.06^{+0.52}_{-0.32})^{-6}$ & 0.64 [13] \\ \\
North 4418 & $2.00^{+*}_{-0.37}$ & $0.98^{+0.26}_{-0.42}$ & $2.5^{+2.6}_{-1.3}$ &
$(7.3^{+1.2}_{-0.8})^{-7}$ & 1.05 [19] \\ \\
South 3726 & $2.00^{+*}_{-0.66}$ & $1.52^{+*}_{-0.81}$ & $1.5^{+4.3}_{-1.0}$ &
$(3.3^{+4.4}_{-0.8})^{-7}$ & 1.29 [6] \\ \\
South 4418 & $2.00^{+*}_{-0.47}$ & $0.68^{+0.64}_{-0.24}$ & $17^{+94}_{-16}$ &
$(4.2^{+1.0}_{-0.6})^{-7}$ & 1.43 [11]
\enddata
\end{deluxetable}


\clearpage

\begin{thebibliography}{}

\bibitem[Arnaud(1996)]{arnaud1996}
Arnaud, K. A. 1996, in ASP Conf. Ser.:Astronomical Data Analysis
Software and Systems V, ed. Jacoby, G. H. \& Barnes, J., 101, 17:
http://heasarc.gsfc.nasa.gov/xanadu/xspec

\bibitem[Bhardwaj \& Gladstone(2000)]{bhardwaj2000}
Bhardwaj, A. \& Gladstone, G.~R.\ 2000, Rev. Geophys., 38, 295

\bibitem[Bhardwaj et~al.(2005)]{bhardwaj2005}
Bhardwaj, A., et~al.\ 2005, \grl, 32, 3S08

\bibitem[Bhardwaj et~al.(2006)]{bhardwaj2006}
Bhardwaj, A., et~al.\ 2006, \jgr, 111, 11225

\bibitem[Bhardwaj et~al.(2007)]{bhardwaj2007}
Bhardwaj, A., et~al.\ 2007, \planss, 55, 1135

\bibitem[Bodewits et al.(2007)]{bodewits2007} Bodewits, D., 
Christian, D. J., Torney, M., Dryer, M., Lisse, C. M., Dennerl, K., 
Zurbuchen, T. H., Wolk, S. J., Tielens, A. G. G. M., \&  Hoekstra, R.
\ 2007, \aap, 469, 1183

\bibitem[Branduardi-Raymont et~al.(2004)]{br2004}
Branduardi-Raymont, G., et~al.\ 2004, \aap, 424, 331

\bibitem[Branduardi-Raymont et~al.(2007a)]{br2007a}
Branduardi-Raymont, G., et~al.\ 2007a, \aap, 463, 761

\bibitem[Branduardi-Raymont et~al.(2007b)]{br2007b}
Branduardi-Raymont, G., et~al.\ 2007b, \planss, 55, 1126

\bibitem[Branduardi-Raymont et~al.(2008)]{br2008}
Branduardi-Raymont, G., et~al.\ 2008, \jgr, 113, 2202

\bibitem[Cravens et~al.(1995)]{cravens1995}
Cravens, T.~E., Howell, E., Waite, J.~H., Jr. \& Gladstone, G.~R.\ 1995,
\jgr, 100, 17153

\bibitem[Cravens et~al.(2003)]{cravens2003}
Cravens, T.~E., et~al.\ 2003, \jgr, 108, 1465

\bibitem[Cravens et~al.(2006)]{cravens2006}
Cravens, T.~E., et~al.\ 2006, \jgr, 111, 7308

\bibitem[Elsner et~al.(2005)]{elsner2005}
Elsner, R.~F., et~al.\ 2005, \jgr, 110, 1207

\bibitem[Gladstone et~al.(1998)]{gladstone1998}
Gladstone, G.~R., Waite, J.~H., Jr., \& Lewis, W.~S.\ 1998,
\jgr, 103, 20083

\bibitem[Gladstone et~al.(2002)]{gladstone2002}
Gladstone, G.~R., et~al.\ 2002, \nat, 415, 1000

\bibitem[Horanyi et~al.(1988)]{horanyi1988}
Horanyi, M., Cravens, T.~E., \& Waite, J.~H., Jr.\ 1988, \jgr, 93, 7251

\bibitem[Johnson et~al.(2002)]{johnson2002}
Johnson, W.~R., Savukov, I.~M., Safronova, U.~I., \& Dalgarno, A.\ 2002, \apj,
141, 543

\bibitem[Kharchenko et~al.(1998)]{kharchenko1998}
Kharchenko, V., Liu, W., \& Dalgarno, A.\ 1998, \jgr, 103, 26687

\bibitem[Kharchenko \& Dalgarno(2000)]{kharchenko2000}
Kharchenko, V. \& Dalgarno, A.\ 2000, \jgr, 105, 18351

\bibitem[Kharchenko et~al.(2006)]{kharchenko2006}
Kharchenko, V., Dalgarno, A., Schultz, D.~R., \& Stancil, P.~C.\ 2006,
\grl, 33, 11105

\bibitem[Kharchenko et~al.(2008)]{kharchenko2008}
Kharchenko, V., Bhardwaj, A., Dalgarno, A., Schultz, D.~R., \& Stancil,
P.~C.\ 2008, \jgr, 133, 8229

\bibitem[Kingston et~al.(2002)]{kingston2002}
Kingston, A.~E, Norrington, P.~H., \& Boone, A.~W.\ 2002, J. Phys. B., 35,
4077
	
\bibitem[Lisse et al.(2001)]{lisse2001} Lisse, C. M., Christian, D. J., 
Dennerl, K., Meech, K. J., Petre, R., Weaver, H. A., \&  Wolk, S. J.
\ 2001, Science 292, 5220

\bibitem[Lisse et al.(2005)]{lisse2005} Lisse, C. M., Christian, D. J., 
Dennerl, K., Wolk, S. J., Bodewits, D., Hoekstra, R., Combi, M. R., 
M\"akinen, T., Dryer, M., Fry, C. D., \&  Weaver, H. \ 2005, \apj, 635,
1329

\bibitem[Lisse et al.(2007)]{lisse2007} Lisse, C. M., Dennerl, K., 
Christian, D. J., Wolk, S. J., Bodewits, D., Zurbuchen, T. H., 
Hansen, K. C., Hoekstra, R., Combi, M., Fry, C. D., Dryer, M., 
M\"akinen, T., \&  Sun, W. \ 2007, Icarus, 190, 391

\bibitem[Liu \& Schultz(1999)]{liu1999}
Liu, W. \& Schultz, D.~R.\ 1999, \apj, 526, 538

\bibitem[Mauk et~al.(2003)]{mauk2003}
Mauk, B.~H., Mitchell, D.~G., Krimigis, S.~M., Roelof, E.~C., \& Paranicas,
C.~P.\ 2003, \nat, 421, 920,

\bibitem[Maurellis et~al.(2000)]{maurellis2000}
Maurellis, A.~N., Cravens, T.~E., Gladstone, G.~R., Waite, J.~H. \& Acton
L.~W.\ 2000, \grl, 27, 1339

\bibitem[Metzger et~al.(1983)]{metzger1983}
Metzger, A.~E., et~al.\ 1983, \jgr, 88, 7731

\bibitem[Nahar(2002)]{nahar2002}
Nahar, S.~N.\ 2002, \aap, 389, 716

\bibitem[Ralchenko et~al.(2008)]{nist}
Ralchenko, Yu., Kramida, A.~E., Reader, J., \& NIST ASD Team\ 2008, NIST
Atomic Spectra Database (version 3.1.5), [Online]. Available:
http://physics.nist.gov/asd3 [2008, December 11]. National Institute of
Standards and Technology, Gaithersburg, Maryland. 

\bibitem[Schultz et~al.(2009)]{schultz2009}
Schultz, D.~R., et~al.\ 2009, Atomic Data and Nuclear Data Tables, in prep.
	
\bibitem[Schwadron \& Cravens(2000)]{schwadron2000} Schwadron, N. A.
\& Cravens, T. E. \ 2000, \apj, 544, 558

\bibitem[van~Hoof(1999)]{all204}
van~Hoof, P.\ 1999, Atomic Line List (version 2.04), [online]. Available:
http://www.pa.uky.edu/$\sim$peter/atomic/ . Department of Physics and Astronomy,
University of Kentucky, Lexington, Kentucky.

\bibitem[von Steiger et~al.(2000)]{vonS2000}
von Steiger, R. et al. 2000, \jgr, 105, 27217

\bibitem[Waite et~al.(1994)]{waite1994}
Waite, J.~H., Jr., et~al.\ 1994, \jgr, 99, 14799

\end{thebibliography}
\end{document}